\documentclass[showpacs,aps,pra,twocolumn,superscriptaddress,longbibliography]{revtex4-2}
\usepackage{geometry}
\usepackage{amsfonts}
\usepackage{graphicx}
\usepackage{breakcites}
\usepackage{amsmath}
\usepackage{hyperref}
\usepackage{cleveref}
\usepackage{mathtools}
\usepackage{physics}
\usepackage{color}
\usepackage{microtype}
\usepackage[utf8]{inputenc}
\usepackage{comment}

\begin{document}
\title{General memory kernels and further corrections to the variational path integral approach for the Bogoliubov-Fr\"{o}hlich Hamiltonian}

\author{T. Ichmoukhamedov}
\affiliation{TQC, Universiteit Antwerpen, Universiteitsplein 1, 2610 Antwerpen, Belgium}
\author{J. Tempere}
\affiliation{TQC, Universiteit Antwerpen, Universiteitsplein 1, 2610 Antwerpen, Belgium}
\date{\today}

\begin{abstract}
The celebrated variational path integral approach to the polaron problem shows remarkable discrepancies with diagrammatic Monte Carlo for the Bogoliubov-Fr\"{o}hlich Hamiltonian which describes an impurity weakly coupled to a Bose condensed atomic gas.
It has been shown both by a renormalization group approach and by the method of  correlated Gaussian wave functions that the model has a subtle UV divergence caused by quantum fluctuations, which are not captured within Feynman's approach. In this work we address the issues with Feynman's approach and show that by extending the model action to a more general form, and by considering higher order corrections beyond the Jensen-Feynman inequality, a good agreement with diagrammatic Monte Carlo can be obtained. 
\end{abstract}


\maketitle

\section{INTRODUCTION}\label{introduction}

Feynman's variational path integral approach \cite{Feynman1955} has since its inception been regarded as the semi-analytical tool of choice to study the Fr\"{o}hlich Hamiltonian \cite{Frohlich}. 
The Fr\"{o}hlich Hamiltonian describes an electron interacting with a bath of phonons in a crystal lattice and is used to study the large polaron in solids \cite{DevreeseReview}. 
In the path integral representation the phonon degrees of freedom can be integrated out exactly, yielding an \textit{effective action} $\mathcal{S}_{\textrm{eff}}$ where the electron is interacting with itself at previous times. 
A variational upper bound for the free energy corresponding to the
effective action can be found in terms of a simpler \textit{model action} $\mathcal{S}_{0}$. 
Feynman's original proposal for the model action \cite{Feynman1955,FeynmanStatisticalMechanics} consists out of a coupled harmonic oscillator with two variational parameters, where one of the particles has been integrated out to simulate the memory effects. 

Regardless of its simplicity, Feynman's approach shows remarkable agreement with computationally demanding diagrammatic Monte Carlo (DiagMC) calculations for the optical \cite{Mishchenko} and the acoustic \cite{Vlietinck2015} polaron. 
In particular, at weak and strong coupling the approach reduces to respectively the coherent state Lee-Low-Pines method \cite{LLPoriginal} and the strong-coupling Landau-Pekar ansatz \cite{pekar1946} and has for this reason also been called \textit{the all-coupling approach} for the Fr\"{o}hlich model. 
Two distinct ways to improve on Feynman's original proposal can be found in the literature. 
First, the model action can be generalized to the best quadratic action functional \cite{Adamkowski,Rosenfelder}, which yields an improvement to Feynman's result for the ground state energy below $0.15 \%$. 
Second, corrections beyond the first order variational expansion can be made \cite{Marshall,SecondOrderCorrection} and yield improvements on Feynman's result below $1.6 \%$. 
These results confirm the astounding accuracy of the simple coupled oscillator model for the ground state energy of the solid state polaron. However, it should be noted that for the dynamical response of the system, an application of the best quadratic action to the optical polaron yields much larger improvements already at weak coupling \cite{Sels2016}. This indicates that some caution should be taken when extrapolating the variational results beyond the ground state energy. 

More recently, a number of experiments have observed the existence of Bose polarons \cite{Hu,Jorgensen,Zwierlein,Ardila2019,Skou2021} where impurities immersed in a Bose-Einstein condensate exhibit quasiparticle properties. 
An impurity with mass $m$, described by first quantization operators $\hat{\mathbf{r}}$ and $\hat{\mathbf{p}}$, couples to the excitations of the Bose gas described in second quantization by the operators $\hat{\alpha}_{\mathbf{k}}^{\dagger}$ and $\hat{\alpha}_{\mathbf{k}}$. 
This closely resembles the solid state polaron and for this reason, at weak coupling,  within the Bogoliubov approximation, and neglecting interactions between impurities, the Bose polaron can be described by the Bogoliubov-Fr\"{o}hlich Hamiltonian \cite{Sacha,Tempere2009Feynman}:
\begin{align}
\hat{H} = & \frac{ \hat{\mathbf{p}}^2}{2m}  +\sum_{\mathbf{k }} \hbar \omega_{\mathbf{k}} \hat{\alpha}^{\dagger}_{\mathbf{k}} \hat{\alpha}_{\mathbf{k}} \nonumber \\
& +\frac{\sqrt{N_0}g_{ib}}{V} \sum_{\mathbf{k  }} V_{\mathbf{k}}  e^{i \mathbf{k} \cdot \hat{\mathbf{r}}}  \left( \hat{\alpha}^{\dagger}_{\mathbf{-k}} + \hat{\alpha}_{\mathbf{k}} \right). \label{BFH}
\end{align}
Here, $N_0$ is the number of condensed bosons, $g_{ib}$ is the contact interaction coupling parameter between the impurity and the bosons, and $V$ is a finite volume in which the Bose gas exists. 

Expression (\ref{BFH}) closely resembles the Fr\"{o}hlich Hamiltonian, and the only difference lies in the functional form of the excitation spectrum $\omega_{\mathbf{k}} $ and interaction amplitude $V_{\mathbf{k}}$:
\begin{align}
& \hbar \omega_{\mathbf{k}}  = \sqrt{\frac{\hbar^2 k^2}{2m_b} \left( \frac{ \hbar^2 k^2}{2m_b} + 2 g_{bb} n_0 \right)},  \label{omegak} \\
& V_{\mathbf{k}} = \left(  \frac{ \frac{ \hbar^2 k^2}{2m_b}}{ \frac{ \hbar^2 k^2}{2m_b}+ 2 g_{bb} n_0 } \right)^{1/4}.
\end{align}
Here, $m_b$, $n_0$ and $g_{bb}$ are respectively the boson mass, density and intraspecies interaction strength. 
Whereas in the original Fr\"{o}hlich Hamiltonian, $\omega_{\mathbf{k}} $ is the constant frequency of longitudinal optical phonons and $V_{\mathbf{k}}$ tends to zero at large momenta, in the Bogoliubov-Fr\"{o}hlich Hamiltonian the coupling amplitude remains finite and the excitation spectrum becomes particle-like. 
This seemingly innocuous change has dramatic consequences for the UV behavior of the model, which from a mathematical physics point of view does not fall into any class of UV divergencies previously encountered in Fr\"{o}hlich-like Hamiltonians \cite{Lampart}. 

It is important to emphasize that beyond weak coupling between the impurity and the gas, the physics of the Bose polaron is not accurately captured by the Bogoliubov-Fr\"{o}hlich Hamiltonian (\ref{BFH}). 
At stronger interactions the Bogoliubov approximation appears to suffer from an instability for attractive polarons \cite{DemlerRG2017,Ichmoukhamedov2019} which was explored in great detail in a more recent study \cite{Schmidt2021}. 
In addition, inclusion of higher order interactions on top of the lowest-order Fr\"{o}hlich coupling term have been considered and were shown to be of importance \cite{SPRath,SchadilovaDynamics,DemlerRG2017,Ichmoukhamedov2019}. 
Finally, the Bogoliubov-Fr\"{o}hlich model does not capture Efimov physics that also play a role in a complete description \cite{LevinsenEfimov,SunEfimov, YoshidaEfimov,Christianen2021}. 
In the rest of this work, we will solely focus on a discussion of the Bogoliubov-Fr\"{o}hlich model (\ref{BFH}) with repulsive effective interactions $g_{ib}$. The discussion will also concern results at stronger coupling and at large momentum cutoff, which are not to be interpreted as a prediction for the Bose polaron in that regime, but rather as a testing ground for corrections to the path-integral approach.  

The Bogoliubov-Fr\"{o}hlich Hamiltonian can also be studied within Feynman's variational approach \cite{Feynman1955,Tempere2009Feynman}, where after the phonons are integrated out, the partition function of the polaron is expressed as a single-particle path integral,
 \begin{equation}
 \mathcal{Z} = \int \mathcal{D} \mathbf{r} ~e^{-\mathcal{S}_{\textrm{eff}}[\mathbf{r}]/ \hbar}. 
 \label{partfunc}
 \end{equation}
The effective action in this path integral contains non-quadratic interactions, which are in addition non-local in time:
\begin{align}
\mathcal{S}_{\textrm{eff}} &=  \int_{0}^{\hbar \beta} \frac{m \dot{\mathbf{r}}^2}{2} d\tau  - \frac{1}{V} \sum_{\mathbf{k}} \frac{g_{ib}^2 n_0}{2 \hbar } \ V_{\mathbf{k}}^2 \nonumber  \\
& \times \int \limits_{0}^{\hbar \beta}  d\tau \int \limits_{0}^{\hbar \beta}  d\sigma   ~\mathcal{G}_{\mathbf{k}} \left(\tau-\sigma \right)  e^{i \mathbf{k} \cdot \left[ \mathbf{r}(\tau)-\mathbf{r}(\sigma) \right]},
\label{effectiveaction}
\end{align}
so that the partition function (\ref{partfunc}) cannot be obtained analytically. 
In Expression~(\ref{effectiveaction}), $\beta = \left( k_B T \right)^{-1}$ is the inverse temperature with Boltzmann factor $k_B$ and
\begin{equation}
\mathcal{G}_{\mathbf{k}}(u) = \frac{\cosh \left[ \omega_{\mathbf{k}} \left( |u| - \hbar \beta/2 \right)\right]}{\sinh \left( \omega_{\mathbf{k}} \hbar \beta/2 \right)}
\end{equation}
is the Green's function of the excitations. For any model action $\mathcal{S}_0$, the Jensen-Feynman inequality provides an upper bound to the free energy $F$ of (\ref{partfunc}):
\begin{equation}
F \leq F_0 + \frac{1}{\hbar \beta} \expval{\mathcal{S}_{\textrm{eff}}- \mathcal{S}_0},
\label{Jensen}
\end{equation}
where $F_0$ is the free energy of the model action and the expectation value in (\ref{Jensen}) is taken with respect to the model system as well. 

The Bogoliubov-Fr\"{o}hlich model has been studied within this approach \cite{Tempere2009Feynman}, where Feynman's original $\mathcal{S}_0$ has been used \cite{FeynmanStatisticalMechanics}. 
Just as is the case for the solid state polaron, the variational energy contains the coherent state result in its weak coupling limit \cite{Nakano}, and hence was expected to work well for this Hamiltonian. 
However, not long afterwards, very unexpectedly large discrepancies between the theory and rigorous DiagMC calculations \cite{Vlietinck2015} have been observed. 
In addition to the well known linear UV divergence in the momentum integrals, associated with using contact interactions, a novel logarithmic UV divergence was argued to be present in the DiagMC study \cite{GrusdtRGFrohlich}. 
The logarithmic UV behavior is completely absent in the variational approach \cite{Tempere2009Feynman}, which is indicative of new physics that is not captured within the approach. Quantum Monte Carlo methods for the Bogoliubov-Fr\"{o}hlich model have also been recently used to study the impurity tunneling problem \cite{Popova}.

In an impressive series of papers by  Grusdt \textit{et al.}, employing a renormalization group (RG) theory \cite{GrusdtRGFrohlich, 2DRGtheory,AllCouplingRGFrohlich,GrusdtReview}, and by Shchadilova \textit{et al.} employing correlated Gaussian wavefunctions (CGW) \cite{ShchadilovaGaussianFrohlich}, the Bogoliubov-Fr\"{o}hlich model has been studied in great detail. 
The authors show that the ground state of the Bogoliubov-Fr\"{o}hlich Hamiltonian contains entangled phonon modes at different energies \cite{ShchadilovaGaussianFrohlich}, and that adequately capturing quantum fluctuations in the RG or CGW approaches gives rise to the logarithmic UV divergence of the ground state energy in the momentum cutoff that can also be observed in DiagMC calculations \cite{Vlietinck2015}.
The momentum cutoff $\Lambda$ therefore plays an important role in the problem, dictating the importance of quantum fluctuations. 
In particular, at large cutoff in the intermediate coupling regime $\alpha \approx 1$, the phonons are argued to be strongly correlated forming the most challenging theoretical regime. 
When compared at small cutoff values and strong coupling, the Jensen-Feynman approach performs better than perturbative RG \cite{GrusdtRGFrohlich} or CGW \cite{ShchadilovaGaussianFrohlich}, and is in good agreement with DiagMC \cite{Vlietinck2015}. 
However, when $\Lambda$ is large, Feynman's approach fails to completely capture the quantum fluctuations and the other approaches provide a far more accurate description, in particular at weak and intermediate coupling. 
More recently, the perturbative RG approach has been extended to also work well at strong coupling \cite{AllCouplingRGFrohlich} lifting it to the status of an all-coupling approach. 
On the other hand, CGW \cite{ShchadilovaGaussianFrohlich} works well at weak to intermediate coupling but shows significant discrepancies with DiagMC towards strong coupling. 
It is curious to note that in the study of the original Fr\"{o}hlich model, Feynman's approach is celebrated precisely for its ability to capture quantum fluctuations when compared with adiabatic density functional theory \cite{DFT2018}, which only emphasizes the elusiveness of the Bogoliubov-Fr\"{o}hlich model in comparison with its solid state counterpart. 
Although the Bogoliubov-Fr\"{o}hlich Hamiltonian is now better understood, nevertheless the question remains as to why Feynman's approach fails or how it can be improved. 
This can be of interest purely from a mathematical point of view \cite{Lampart}, or as a first step towards future applications to multiple particles in this model or extended Fr\"{o}hlich Hamiltonians \cite{SchadilovaDynamics,Ichmoukhamedov2019}. 
In addition, the method employed in this work will illustrate the utility of general memory kernels for variational applications which we have also considered in another context \cite{Ichmoukhamedov2021}.  
The central goal of this paper is therefore to use the Bogoliubov-Fr\"{o}hlich model as an illustration of the importance of further corrections to the path integral method when applied to polaronic models where quantum fluctuations cause additional UV divergences.

 Note that in \cite{GrusdtRGFrohlich}, a regularization procedure of this UV divergence is proposed through effective mass corrections to the mean field impurity-condensate interactions term $g_{ib}n_0$, which we have not included in the Hamiltonian (\ref{BFH}). Here, we will not be concerned with this regularization since the goal is specifically to discuss the mechanism of appearance of this UV behavior in Feynman's approach. Moreover, for an accurate comparison with realistic experiments the cutoff $\Lambda$ should be related to either the inverse van der Waals length of the atomic potential \cite{Tempere2009Feynman,DemlerRG2017} or the first Efimov resonance \cite{Christianen2021}. For the current system the former value corresponds to $\Lambda \approx 200 \xi^{-1}$, and in what follows figures will be presented with results up to $\Lambda \approx 4000 \xi^{-1}$. For this reason, in addition to the Bogoliubov-Fr\"{o}hlich model being only valid at weak coupling, we emphasize that the results in their current form are not suitable for direct comparison with experiment.    

In this paper we show that two modifications bring significant improvements to Feynman's approach for the Bogoliubov-Fr\"{o}hlich Hamiltonian. 
In Sec.~\ref{BestQuadraticAction} we consider the best quadratic action functional \cite{Rosenfelder} as the model action for this system. 
While this correction is extremely small for the energy of the Fr\"{o}hlich model, we show that the largest correction to the ground state energy of the Bogoliubov-Fr\"{o}hlich model is obtained in this step. 
We obtain the optimal memory kernel which also provides insights as to why Feynman's original model fails. 
At strong coupling the results show good agreement with DiagMC, but near the challenging intermediate coupling regime some noticeable discrepancy remains. 

To obtain further corrections for the intermediate regime, in Sec.~\ref{SecondOrder} we derive an expression for the correction from the second-order cumulant expansion of the partition function. 
This correction has been shown to be small \cite{Marshall,SecondOrderCorrection} in the Fr\"{o}hlich model, but turns out to be appreciable for the Bogoliubov-Fr\"{o}hlich model. 
Combining the two aforementioned improvements, we retrieve the logarithmic divergence of the model and find excellent agreement with DiagMC in the intermediate regime.

\section{Quadratic action with a general memory kernel}\label{BestQuadraticAction}

The derivation presented here has been performed for the Fr\"{o}hlich model in \cite{Adamkowski}, and further addressed in \cite{Rosenfelder,Sels2016}. 
Contrary to the derivation in \cite{Adamkowski} that we will follow here, the momentum integrals cannot be analytically performed in the Bogoliubov-Fr\"{o}hlich model and hence we briefly review the derivation, now applied to (\ref{BFH}). 
The central quantity in this section will be the model action functional (working in units of $\hbar=1$ from now on):
\begin{align}
\mathcal{S}_0 & =  \frac{m}{2} \int_0^{\beta} \dot{\mathbf{r}}^2 d\tau \nonumber  \\
& +  \frac{ m}{2 \beta} \int_0^\beta \int_0^\beta d\tau d\sigma ~ x(\tau-\sigma) \mathbf{r}(\tau) \cdot \mathbf{r}(\sigma),
\label{S0}
\end{align}
where $x(\tau-\sigma)$ is a general memory kernel with greater freedom than the commonly used Feynman model action. 
Note that introducing an additional $\beta$ in the denominator of the second term in (\ref{S0})  will prove to be convenient further on. 
Following \cite{Adamkowski}, we make the restriction to $\beta$-periodic functions $x(\beta-\tau)=x(\tau)$ and in addition assume $\int_0^\beta x(\tau) d\tau =0$. 
While the first assumption is necessary for the derivation, the second could in principle be relaxed \cite{Adamkowski}. 
The goal is to find an expression for the variational free energy (\ref{Jensen}) as a functional of the memory kernel $x(\tau-\sigma)$:
\begin{equation}
F_{v}[x] = F_0 + \frac{1}{\beta} \expval{\mathcal{S}_{\textrm{eff}}- \mathcal{S}_0 }.
\label{Fv}
\end{equation} 
Since the action functional is quadratic in the impurity degree of freedom, exact expressions for all quantities in (\ref{Fv}) can be obtained. 
In what follows we summarize the steps in \cite{Adamkowski}, now applied to the Bogoliubov-Fr\"{o}hlich model. 
In principle, all expectation values of analytic functions of $\mathbf{r}(\tau)$ can 
be computed via a generating function, which satisfies the following identity
for any vector function $\mathbf{g}(\tau)$ (in three dimensions):
\begin{align}
 & \expval{ \exp \left( \int \limits_0^{\beta} \mathbf{g}(\tau) \cdot \mathbf{r}(\tau) d\tau \right) }   \nonumber \\
& =  \exp \left( \frac{1}{6} \int \limits_0^{\beta} \int \limits_0^{\beta}   \expval{ \mathbf{r}(\tau) \cdot \mathbf{r}(\sigma) } \mathbf{g}(\tau) \cdot \mathbf{g}(\sigma)  d\tau d\sigma \right)   . \label{generatingfunction}
\end{align}
The property of $\beta$-periodicity allows one to decompose the memory kernel in Fourier space $x(u) = \sum_{n=-\infty}^{\infty} x_n e^{i \nu_n u}$ with Matsubara frequencies $\nu_n= 2 \pi n/ \beta$. 
The covariance in expression (\ref{generatingfunction}) is nothing else than the Green's function of the corresponding classical equation of motion, as commonly encountered in introductory quantum field theory \cite{zee2010quantum}. 
Here, it can also be obtained in first quantization:
\begin{equation}
\expval{ \mathbf{r}(\tau) \cdot \mathbf{r}(\sigma) } = \frac{6}{m \beta} \sum_{n=1}^{\infty}  \frac{\cos \left[ \nu_n (\tau-\sigma)\right]}{\nu_n^2 + x_n} .
\label{covariance}
\end{equation}

If an auxiliary parameter $\lambda$ is introduced in the action functional (\ref{S0}) as a scaling factor to the memory kernel $x(\tau-\sigma) \rightarrow \lambda x(\tau-\sigma)$, the partition function $\mathcal{Z}$ and free energy $F_0$ obtain a $\lambda$ dependence, and it can be readily shown that:
\begin{equation}
\frac{\partial F_0^{(\lambda)} }{\partial \lambda} = \frac{m}{2 \beta^2} \int \limits_0^\beta \int \limits_0^\beta x(\tau-\sigma) \expval{ \mathbf{r}(\tau) \cdot \mathbf{r}(\sigma) }_{\lambda}  d\tau d\sigma .
\label{dF}
\end{equation}
The subscript $\lambda$ indicates that $x_n$ has been
scaled to $\lambda x_n$ in the covariance (\ref{covariance}). Expression (\ref{dF}) can now be integrated over $\lambda$ to obtain the free energy of the model system:
\begin{align}
F_0^{(\lambda)}  =& - \frac{1}{\beta} \log \left[ \left( \frac{m}{2 \pi \beta} \right)^{3/2}  V \right]  \nonumber \\
&+ \frac{3}{\beta} \sum_{n=1}^{\infty} \log( 1 + \frac{\lambda x_n}{\nu_n^2} ).
\label{F_0}
\end{align}

The kinetic energy contributions to the action functionals cancel in the second term of (\ref{Fv}) and hence it is useful to redefine $\mathcal{\tilde{S}}_0$ and $ \mathcal{\tilde{S}}_{\textrm{eff}}$, where the absence of the kinetic energy terms is emphasized by the tilde. 
By once again introducing the auxiliary variable and taking the derivative of the partition function with respect to $\lambda$, one can show:
\begin{equation}
\frac{1}{\beta} \expval{\mathcal{\tilde{S}}_0} = \left. \frac{\partial F_0^{(\lambda)} }{\partial \lambda} \right|_{\lambda=1} =  \frac{3}{\beta} \sum_{n=1}^{\infty} \frac{x_n}{x_n + \nu_n^2}.
\end{equation}
The generating function result (\ref{generatingfunction}) also immediately yields the expectation value of the effective action (\ref{effectiveaction}). 
Note that the covariance (\ref{covariance}) only depends on the time difference $|\tau-\sigma|$ and is in addition $\beta$-periodic. 
In the limit of zero temperature $\beta \rightarrow \infty$ this simplifies the double time integral from (\ref{effectiveaction}) to:
\begin{equation}
\frac{1}{\beta} \expval{\tilde{S}_\textrm{eff}}= - \frac{g_{ib}^2 n_0}{V} \sum_{\mathbf{k}} V_{\mathbf{k}}^2 \int \limits_0^{\beta/2} \mathcal{G}_{\mathbf{k}}(u) \mathcal{F}_{\mathbf{k}}(u)  du  \label{expSeff}
\end{equation}
where:
\begin{equation}
 \mathcal{F}_{\mathbf{k}}(u) = \exp \left( - \frac{2k^2}{m \beta} \sum_{n=1}^{\infty} \frac{1- \cos(\nu_n u) }{x_n + \nu_n^2} \right).
 \label{Fdiscrete}
\end{equation}
In the $\beta \rightarrow \infty$ limit the Matsubara summations in the previous expressions are transformed into frequency integrals, where the coefficients $x_n$ become the Fourier transform $x(\nu)$ of the memory kernel (notation not to be confused with the original function): 
\begin{equation}
 \mathcal{F}_{\mathbf{k}}(u) = \exp \left( - \frac{k^2}{\pi m} \int \limits_0^{\infty} d\nu \frac{1- \cos(\nu u)}{x(\nu)+ \nu^2} \right).
\end{equation}

Expression (\ref{expSeff}) contains a linear divergence in the momentum summation and is regularized by relating $g_{ib}$ to the s-wave scattering length $a_{ib}$ up to second order in the Lippmann-Schwinger equation in the Bose-polaron mean-field energy $g_{ib} n_0$ \cite{Tempere2009Feynman}. 
This regularization eventually comes down to simply using the lowest-order expression for $g_{ib}= 2 \pi  \hbar^2 a_{ib} / \mu $, where $\mu^{-1}= m^{-1} + m_b^{-1}$ is the reduced impurity-boson mass, but now subtracting the divergent behavior from (\ref{expSeff}). 
For the Bose intraspecies interaction a lowest order expression $g_{bb}=4\pi \hbar^2 a_{bb}/m_{b}$ is sufficient. 
Note that this regularization procedure is commonly performed in systems with contact interactions and this divergence is not related to the phonon entanglement discussed in the introduction. 

\begin{figure*}[!htbp]
\includegraphics[width=0.9\textwidth]{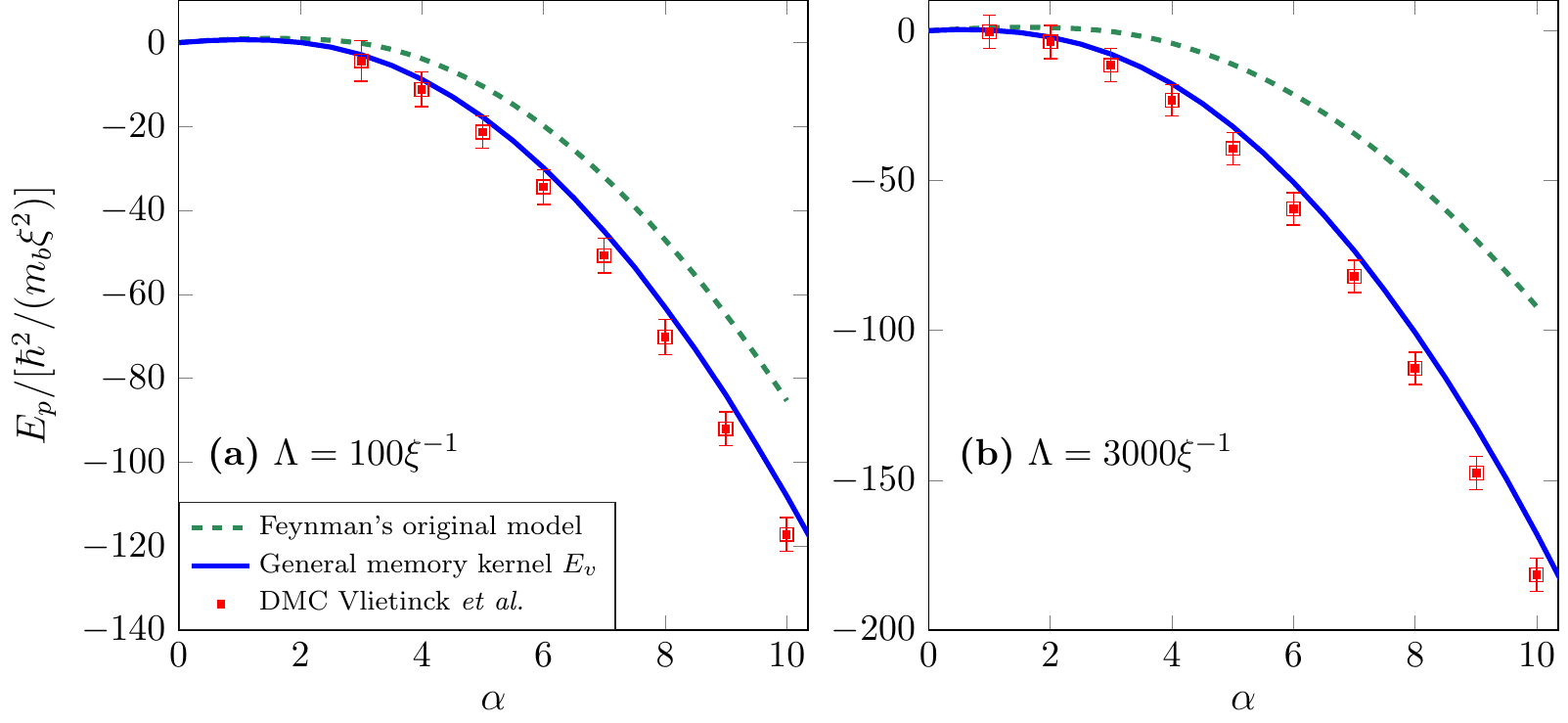}
\caption{The ground state energy at zero temperature ($\beta=200$ used as a cutoff) for (a) moderate $\Lambda=100 \xi^{-1}$ and (b) large $\Lambda=3000 \xi^{-1}$ cutoff values, compared to the results taken from diagrammatic Monte Carlo calculations \cite{Vlietinck2015} (scatter squares). The dashed line shows an application of Feynman's original model \cite{Tempere2009Feynman} to this system, while the solid line is our result obtained with the general quadratic memory kernel. Note that for this comparison the energy scale is defined using the boson mass $m_b=3.8m$.} 
\label{Figure1}
\end{figure*}

Unless specified otherwise, in the rest of the text we will use polaronic units of \cite{Tempere2009Feynman} in terms of the impurity mass $m=1$, the condensate healing length $\xi = \hbar / \sqrt{ 2 m_b g_{bb} n_0}=1$, and the corresponding energy scale $\hbar^2/( m \xi^2)=1$, which also corresponds to setting $\hbar=1$. 
In the rest of this work the mass ratio $m_b=3.8m$ is used for all the figures for comparison with the DiagMC results of \cite{Vlietinck2015}. Note that to facilitate comparison with \cite{Vlietinck2015} where the boson mass $m_b$ was preferred as the mass unit, an appropriate energy rescaling is performed on the figures. 
The dimensionless coupling constant of this model \cite{Tempere2009Feynman} is then given by $\alpha = a_{ib}^2/(a_{bb} \xi)$. 
Combining all of the previous terms, taking the $\beta \rightarrow \infty$ limit, and also taking the volume $V$ to infinity, allows one to write the variational functional as:
\begin{align}
F_v[x] = & \frac{3}{2\pi} \int \limits_0^{\infty} d\nu \left[ \log(1 + \frac{x(\nu)}{\nu^2} ) - \frac{x(\nu)}{x(\nu)+\nu^2}  \right] \nonumber  \\
& - \frac{\alpha}{4\pi \mu^2} \int \limits_0^{\Lambda} dk~k^2 V_{\mathbf{k}}^2 \int \limits_0^{\beta/2} \mathcal{G}_{\mathbf{k}}(u) \mathcal{F}_{\mathbf{k}}(u) du \nonumber \\
&+ \frac{\alpha \Lambda}{2\pi \mu} . 
\label{FvBest}
\end{align}
Here, $\Lambda$ is the finite momentum cutoff discussed in Sec.~(\ref{introduction}), and the final term arises from the contact interaction regularization. 
The functional that minimizes the energy is found by taking the derivative with respect to a discrete Fourier component $\partial_{x_n} F_v=0$ before the continuum limit is taken. 
Once the continuum limit is taken, the following integral equation can be obtained for the memory kernel:
\begin{align}
x(\nu)= & ~ \frac{\alpha}{3\pi \mu^2} \int \limits_0^{\Lambda} dk ~ k^4 V_{\mathbf{k}}^2
\nonumber \\
& \times \int \limits_0^{\beta/2} \mathcal{G}_{\mathbf{k}}(u) \mathcal{F}_{\mathbf{k}}(u) \sin(\frac{\nu u}{2})^2 du. \label{xnu}
\end{align} 
Since $\mathcal{F}_{\mathbf{k}}(u)$ is itself a functional of $x(\nu)$ this equation has to be solved numerically. 
This is done iteratively, starting by substituting the Lee-Low-Pines solution $x(\nu)=0$ into the right-hand side of (\ref{xnu}) and obtaining an improved memory kernel on the left-hand side. 
Depending on $\alpha$, roughly one to ten iterations are needed until the relative increase in the corresponding energy (\ref{FvBest}) becomes less than $1 \%$, which we accept as our final value. 
The next iteration yields further corrections of the order of $0.1 \%$ and can no longer be discerned on the graphs shown in this paper. 
We find that the frequency at which the memory kernel reaches an asymptotic value can become very large. 
For this reason we perform a scaling transformation $\nu= e^{z}-1$ and select $N=1000$ Gauss-Legendre quadrature points on the $z$-grid up to $\nu_{max}=10^8$. 
The iterative improvement (\ref{xnu}) is then performed for each point.

The results are shown in Fig. (\ref{Figure1}) where we compare the ground state energy (\ref{FvBest}) for the optimized memory kernel with DiagMC results from \cite{Vlietinck2015}. 
As already observed 
in \cite{Vlietinck2015,ShchadilovaGaussianFrohlich,GrusdtRGFrohlich},
the original Feynman model yields surprisingly large discrepancies at strong coupling, especially at larger values of the cutoff $\Lambda=3000\xi^{-1}$. 
This indicates that even in the limit of strong coupling, quantum fluctuations are of importance and the adiabatic ansatz, included in Feynman's original model, fails for this system. 
We can see that the result for the best quadratic action functional (\ref{FvBest}) provides significant corrections to Feynman's model and yields a variational bound in good agreement with DiagMC at strong coupling. 
However, as will be shown in Fig.~(\ref{Figure4}) in the next section, in the challenging intermediate coupling regime some discrepancies remain. 
To estimate corrections in this region, in the next section we consider further contributions to the energy beyond the first order variational inequality.

\begin{figure}[!htbp]
\includegraphics[width=0.45\textwidth]{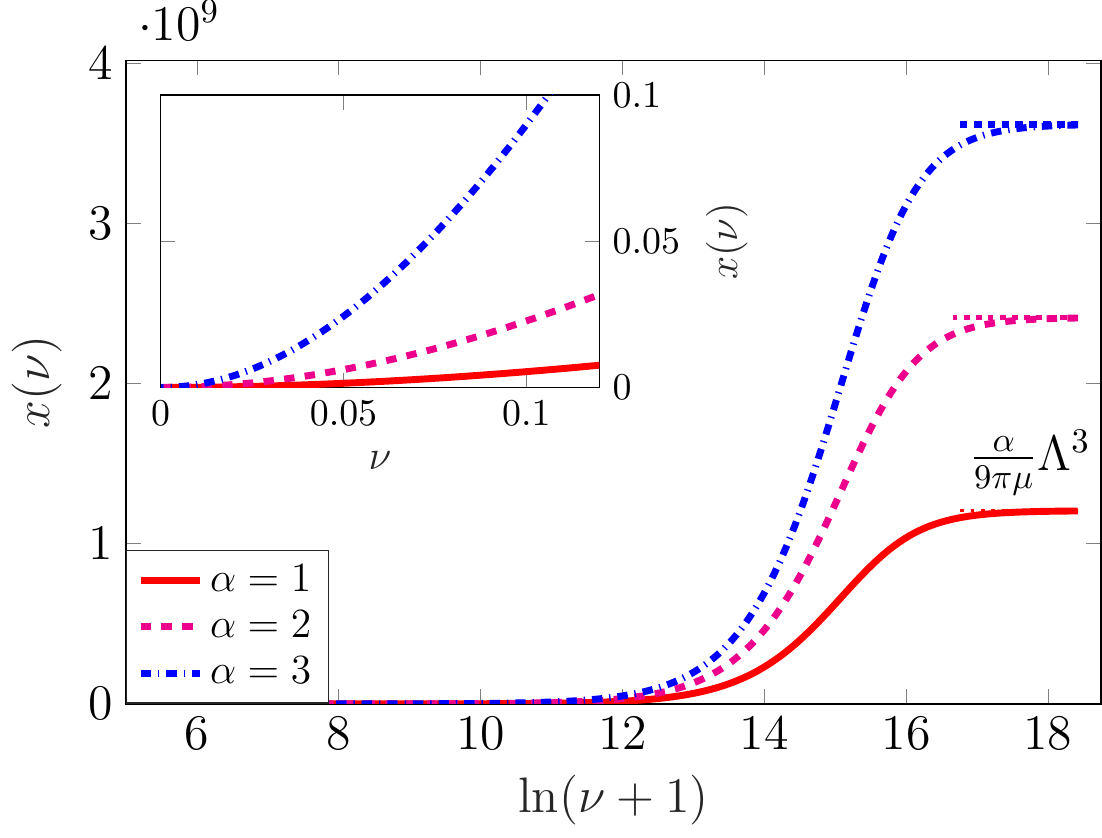}
\caption{ The optimized memory kernel $x(\nu)$ at $\Lambda=3000\xi^{-1}$ (in polaronic units) obtained for the energy plot in Fig (\ref{Figure1}b) at three different coupling strengths $\alpha=1,2,3$. The main plot shows the behavior at large $\nu$ on a logarithmic frequency axis, whereas the inset indicates a quadratic behavior at small frequencies. The dashed horizontal lines represent the analytic $\nu \rightarrow \infty$ limit mentioned in the text.
} 
\label{Figure2}
\end{figure}

Let us also pay some attention to the optimized memory kernel itself. 
In Fig. (\ref{Figure2}) we show the obtained optimized solutions for $x(\nu)$ that lead to the results shown in Fig.~(\ref{Figure1}). 
We can see that the UV limit of $x(\nu)$ agrees with the analytic expression $x(\nu)= \frac{\alpha}{6\pi \mu^2} \left( 2\mu \Lambda^3/3 \right) $. 
This limit can be readily obtained by substituting the mean field guess $x(\nu)=0$ in the RHS of (\ref{xnu}) and then taking the $\nu \rightarrow \infty$ limit. 
Therefore, it appears that in the Bogoliubov-Fr\"{o}hlich model the UV limit of the optimal memory kernel does not converge as $\Lambda \rightarrow \infty$. 
On the other hand, we have checked that the small frequency behavior shown on the inset of Fig.~(\ref{Figure2}) is only very weakly influenced by the cutoff (while it does depend on $\alpha$).
\begin{figure}[t]
\includegraphics[width=0.475\textwidth]{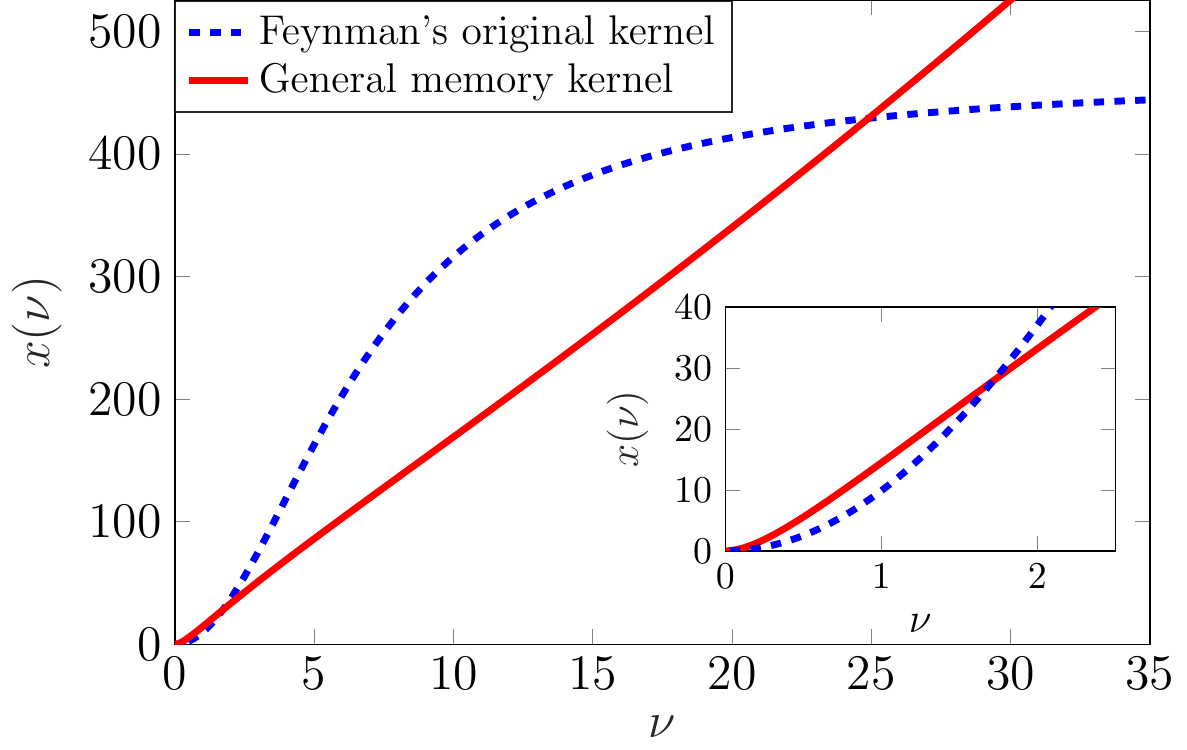}
\caption{A comparison between the general optimized memory kernel $x(\nu)$ and Feynman's original memory kernel. Inset shows the same plot at smaller frequencies. Plots are made at $\Lambda=3000\xi^{-1}$ and $\alpha=5$.} 
\label{Figure3}
\end{figure}

Finally, in Fig.~(\ref{Figure3}) we compare the shape of the optimized memory kernel to Feynman's original model, which is given by $x_{\textrm{Feyn}}(\nu) =  MW^2 \nu^2 / (\nu^2+W^2)$. 
The memory kernel tends to zero quadratically in $\nu$, 
in the $\nu \rightarrow 0$ limit. 
This can be analytically shown in the first iteration by expanding Expression~(\ref{xnu}) to lowest order in $\nu$, and for the optimized solution this behavior is shown in the inset of Fig.~(\ref{Figure2}). 
However, as can be seen in the inset of Fig.~(\ref{Figure3}), 
the quadratic behavior rapidly transitions into an extended linear regime. 
In principle, the memory kernel of Feynman's model system exhibits a similar behavior, it starts as quadratic and then transitions into a linear regime before moving to an asymptotic value. 
The problem however is that Feynman's model has only two variational parameters so that the ranges of the regimes can not all be chosen independently. 
The quadratic behavior at small frequencies is dictated by the parameter $MW^2$ while the transition into the linear regime depends on $W^2$ in the denominator. 
This forces the memory kernel of Feynman's model system to make a compromise and reach its asymptotic value far more quickly than the general solution. 

\section{Second order correction}\label{SecondOrder}

By adding and subtracting the model action functional $\mathcal{S}_0$ in the path integral of the polaron partition function (\ref{partfunc}), the free energy of the system can be exactly written as \cite{Mills}:
 
 \begin{equation}
 F = F_0 - \frac{1}{\beta} \ln( \expval{ e^{-\Delta \mathcal{S}} } ),
 \end{equation}
where $\Delta \mathcal{S} = \mathcal{S}_{\textrm{eff}}-\mathcal{S}_0 $, and the expectation values are taken with respect to the model action. 
The second term can be recognized as the cumulant-generating function of the path integral, which can be expanded as \cite{Mills,Marshall}:

\begin{align}
F&= F_0 + \frac{1}{\beta} \expval{\Delta \mathcal{S}}_0 - \frac{1}{2!} \frac{1}{\beta} \expval{ \left( \Delta \mathcal{S}- \expval{\Delta \mathcal{S}}\right)^2} \nonumber \\
&+ \mathcal{O} \left( \Delta \mathcal{S}^3  \right).
\label{cumulant}
\end{align}
Here, we can recognize the variational free energy in the first two terms, which represent the Jensen-Feynman inequality (\ref{Jensen}) if all the higher order terms are discarded.  

In this section we perturbatively include the second order correction in the cumulant expansion of (\ref{cumulant}). 
This correction has been studied and shown to be small in the Fr\"{o}hlich model \cite{Marshall,SecondOrderCorrection}, but we find it to be non-negligible in the Bogoliubiov-Fr\"{o}hlich model. 
We emphasize that the resulting correction is approximate for two reasons. 
First of all to the best of our knowledge, no general inequalities that include higher orders of the expansion are known \cite{CumulantExpansion} and hence from this point on the variational inequality can be violated. 
Second, the obtained correction is significantly more difficult to compute than the one obtained in \cite{SecondOrderCorrection} due to the fact that the momentum integrals cannot be analytically performed in the Bogoliubov-Fr\"{o}hlich model. 
For this reason a mean-field like approximation to obtain a semi-analytic expression will be made. At weak to intermediate coupling the obtained corrected energy is in excellent agreement with DiagMC and exhibits the exact logarithmic divergence that was observed in \cite{Vlietinck2015}.

In what follows we are strictly interested in the $\beta \rightarrow \infty$ limit. 
For convenience of notation, and to avoid having to write the formal limit everywhere, we will keep the Matsubara summations in their discrete form and still write the factor $\beta$ in e.g.~the integral boundaries. 
Such expressions are to be strictly interpreted on the condition that $\beta$ is very large and will be taken to infinity in the end, on which our derivation relies. The cumulant in the second order correction can be written as:
\begin{align}
 & \frac{1}{2\beta} \expval{ \left( \Delta \mathcal{S}- \expval{\Delta \mathcal{S}}  \right)^2}  =  \frac{1}{2\beta} \left[ \expval{\mathcal{\tilde{S}}_{\textrm{eff}}^2} - \expval{\mathcal{\tilde{S}}_{\textrm{eff}}}^2   \right]  \nonumber \\
  &   +  \frac{1}{2\beta} \left[ \expval{\mathcal{\tilde{S}}_{0}^2} - \expval{\mathcal{\tilde{S}}_{0}}^2   \right]  \nonumber \\
  & -  \frac{1}{ \beta} \left[ \expval{\mathcal{\tilde{S}}_{\textrm{eff}} \mathcal{\tilde{S}}_{0}}  - \expval{\mathcal{\tilde{S}}_{\textrm{eff}}} \expval{\mathcal{\tilde{S}}_{0}}  \right].
  \label{cumulant1}
\end{align}
In Appendix (A) we show that it is convenient to define the following fivefold integral:
\begin{align}
& \sigma_n[x(\nu)] = \frac{1}{(2n+1)!} \left( \frac{\alpha}{4\pi \mu^2} \right)^2   \int \limits_0^{\Lambda} dk  \int \limits_0^{\Lambda} ds   \nonumber \\
& \times k^2 s^2 V_{\mathbf{k}}^2 V_{\mathbf{s}}^2   \int \limits_0^{\beta/2} du_1   \int \limits_0^{\beta/2} du_2 ~ \mathcal{G}_{\mathbf{k}}(u_1) \mathcal{G}_{\mathbf{s}}(u_2)  \nonumber \\
& \times  \mathcal{F}_{\mathbf{k}}(u_1) \mathcal{F}_{\mathbf{s}}(u_2) \int \limits_0^{\beta/2} dz     \left( \frac{ks}{4} \zeta(u_1,u_2,z ) \right)^{2n},
\label{sigmanMain}
\end{align}
where:
\begin{align}
\zeta(u_1,u_2,z) = & \frac{32}{\beta} \sum_{n=1}^{\infty} \left[ \frac{1}{\nu_n^2+x_n}
 \cos(\nu_n z) \right.
 \nonumber \\
 & \left. \times \sin(\frac{\nu_n u_1}{2}) \sin(\frac{\nu_n u_2}{2}) \right].
\label{zetaMain}
\end{align}
For any memory kernel $x(\nu)$, the second order correction (\ref{cumulant1}) can now be written as:
\begin{align}
 & \frac{1}{2\beta} \expval{ \left( \Delta \mathcal{S}- \expval{\Delta \mathcal{S}}  \right)^2} = \sum_{n=2}^{\infty} \sigma_{n}[x(\nu)] \nonumber \\
 & +  \left[ \sigma_{1}[x(\nu)] + \frac{1}{2\beta} \left( \expval{\mathcal{\tilde{S}}_{0}^2} - \expval{\mathcal{\tilde{S}}_{0}}^2   \right)   \right. \nonumber \\
 & \left.   -  \frac{1}{ \beta} \left( \expval{\mathcal{\tilde{S}}_{\textrm{eff}} \mathcal{\tilde{S}}_{0}}  - \expval{\mathcal{\tilde{S}}_{\textrm{eff}}} \expval{\mathcal{\tilde{S}}_{0}}  \right) \right]. 
  \label{cumulant2}
\end{align}
Note that if $x(\nu)=0$ is substituted in the variational free energy (\ref{FvBest}) one obtains the mean-field Lee-Low-Pines result at zero polaron momentum. 
Therefore, for $x(\nu)=0$ we can interpret the result (\ref{cumulant2}) 
as a correction to mean-field theory:
\begin{align}
 & \frac{1}{2\beta} \expval{ \left( \Delta \mathcal{S}- \expval{\Delta \mathcal{S}}  \right)^2}^{(\textrm{MF})} =  \sum_{n=1}^{\infty} \sigma_{n}[0].
  \label{cumulantMF}
\end{align}
Incidentally, the polaron problem mean-field theory corresponds to first-order perturbation theory \cite{GrusdtReview}, and hence (\ref{cumulantMF}) is also nothing else than the second order perturbative correction. Due to the simplification $x(\nu)=0$, the sum in (\ref{cumulantMF}) can be performed. 
However, mean-field theory completely misses the DiagMC polaronic energy in the Bogoliubov-Fr\"{o}hlich Hamiltonian beyond weak coupling \cite{GrusdtRGFrohlich} and hence it is desirable to start from a better point.

\begin{figure*}[!htbp]
\includegraphics[width=0.65\textwidth]{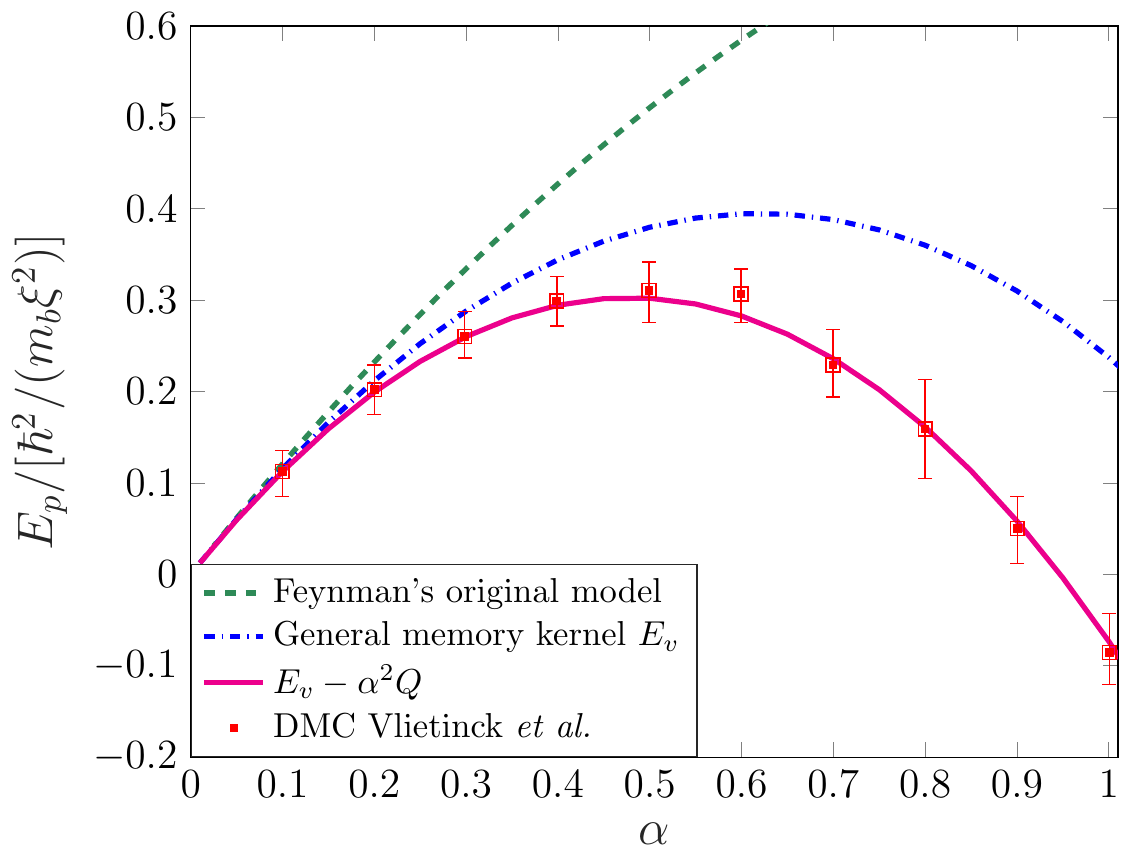}
\caption{The polaronic energy obtained from diagrammatic Monte Carlo 
\cite{Vlietinck2015} (squares with error bars) at $\Lambda=2000 \xi^{-1}$ 
is compared to the result of Feynman's original model action (dashed curve), 
to the result of the general memory kernel method of Sec.~\ref{BestQuadraticAction} (dashdotted curve), 
and to the general memory kernel result including the secord order correction of Sec.~\ref{SecondOrder} (solid curve).
} 
\label{Figure4}
\end{figure*}

\begin{figure}[!htbp]
\includegraphics[width=0.48\textwidth]{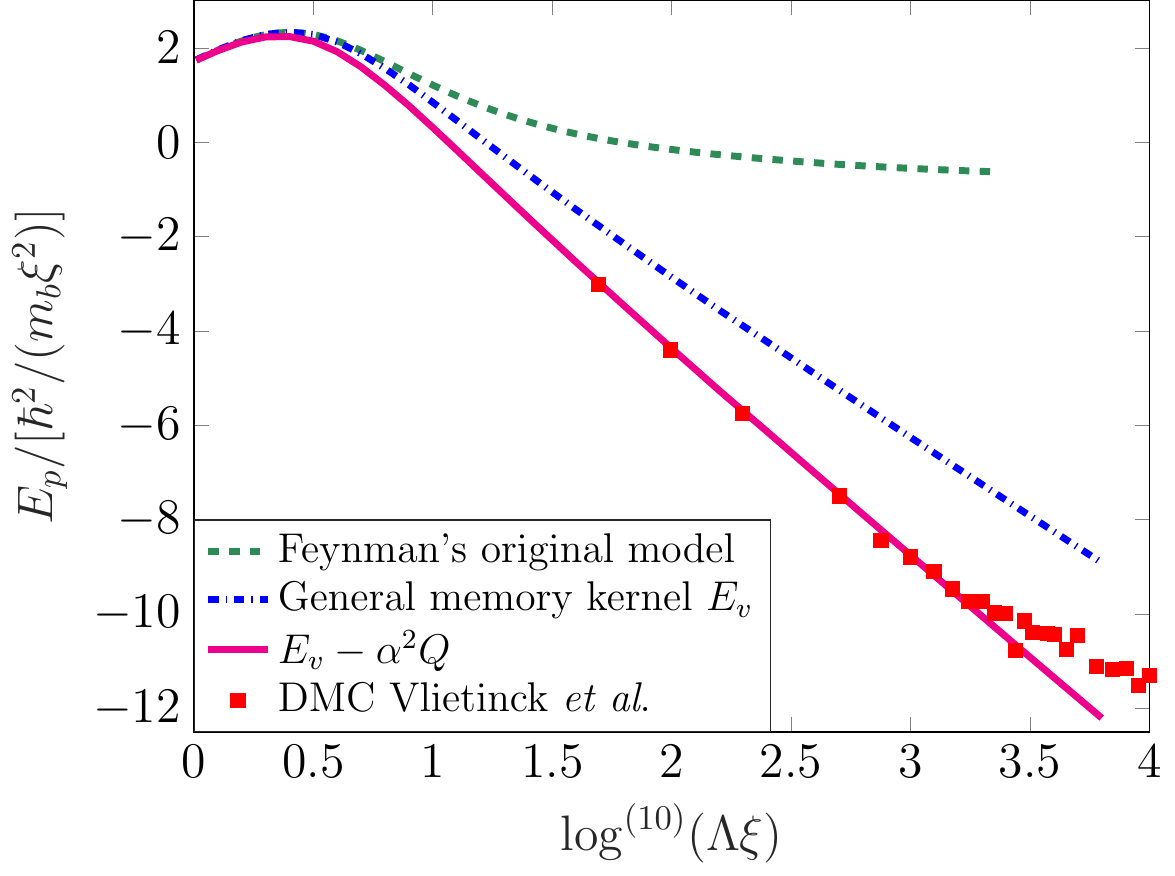}
\caption{The polaronic energy contribution obtained within Feynman's original model (dashed), the general memory kernel method (dashdotted), the corrected energy (solid) and diagrammatic Monte Carlo results taken from \cite{Vlietinck2015} (squares) are plotted for $\alpha=3$ as a function of the cutoff $\Lambda$ on a logarithmic scale.
} 
\label{Figure5}
\end{figure}

Let us now consider the corrections on top of the best quadratic action functional with the optimized memory kernel (\ref{xnu}). 
As shown in Appendix (\ref{AppendixA}), in this case the terms in the square 
brackets in (\ref{cumulant2}) all completely cancel and
the summation starts from $n=2$
\begin{align}
 & \frac{1}{2\beta} \expval{ \left( \Delta \mathcal{S}- \expval{\Delta \mathcal{S}}  \right)^2}^{(\textrm{best})} =  \sum_{n=2}^{\infty} \sigma_{n}[x(\nu)].
  \label{cumulantBest}
\end{align}
Therefore an important feature of expanding around the best quadratic action is to omit the dominant contribution from $\sigma_1$ in the mean-field correction (\ref{cumulantMF}). 
In contrast with previous approaches that have considered the second order correction for the polaron, the momentum integrals in (\ref{sigmanMain}) cannot be performed analytically. 
For any non-trivial memory kernel $x(\nu)$ one is hence left with a fivefold integral, which we have not been able to compute efficiently. 

Let us therefore in spirit of Feynman's approach consider a simple semi-analytic approximation. 
We expand around the best quadratic action in (\ref{cumulant2}) and use this knowledge to cancel the term in the square brackets, but then approximate the remaining contributions at the mean-field level:
\begin{align}
 & \frac{1}{2\beta} \expval{ \left( \Delta \mathcal{S}- \expval{\Delta \mathcal{S}}  \right)^2}^{(\textrm{best})} \approx \sum_{n=2}^{\infty} \sigma_{n}[0].
  \label{cumulantBestMf}
\end{align}
The error of this approximation is roughly estimated by calculating the difference in the first-order term $\sigma_1[x(\nu)]- \sigma_1[0]$, for which the fivefold integral can be easily performed. This difference is obtained in expression (\ref{sigma1c}) in Appendix (\ref{AppendixA}) (which is to be computed in the $\beta \rightarrow \infty$ limit):
\begin{align}
& \sigma_1[x(\nu)] - \sigma_1[0] =  \frac{3}{2 \beta} \sum_{n=1}^{\infty} \left[ \frac{x_n^2}{(\nu_n^2+x_n)^2} - \frac{\tilde{x}_n^2}{\nu_n^4} \right].
\label{SigmaError}
\end{align} 
Here, $x_n$ represents the optimal memory kernel, whereas $\tilde{x}_n$ is the first-order iterative improvement obtained from substituting $x(\nu)=0$ into (\ref{xnu}). 
For a cutoff $\Lambda=2000\xi^{-1}$ the relative error on the correction is of the order of $3 \%$ for $\alpha=0.5$ and of the order $5 \%$ for $\alpha=1$. 
This justifies using the approximation to get an accurate second order correction in the weak to intermediate coupling regime, in particular for Fig.~(\ref{Figure4}). 
This error is however larger at large coupling strengths, and we found an over correction towards energies below DiagMC for $\alpha \gtrsim 5 $ when applied to Fig. (\ref{Figure1}). 
This suggests that even higher order corrections are likely needed to get exactly on DiagMC in that regime, and in what follows we only apply the correction in the weak to intermediate coupling regime.

As shown in Appendix (\ref{AppendixB}), the full corrected energy on top of the minimized $E_v$ from (\ref{FvBest}), with this approximated correction is given by:
\begin{equation}
E=E_v - \alpha^2 Q,
\end{equation}
where:
\begin{align}
Q & =   \sum_{n=2}^{\infty} \frac{1}{(2n+1)} \left( \frac{1}{4\pi \mu^2} \right)^2  \nonumber \\
& \int \limits_0^{\Lambda} dk  \int \limits_0^{\Lambda} ds     V_{\mathbf{k}}^2 V_{\mathbf{s}}^2  \frac{k^{2+2n} s^{2+2n}\left(3 a(k) + a(s) \right)   }{a(k)^2 \left( a(k) + a(s) \right)^{2+2n}}, \label{Q}
\end{align}
with $a(k)= \omega_k + k^2/(2m)$. This double integral is easily performed and the series converges within less than $0.1\%$ after $n=10$. 
If the sum is extended to $n=0$, the series expansion of $x \textrm{arctanh}(x) $ can be recognized here which indicates that the integral could likely be more easily performed in the $\sinh(x)/x$ form in (\ref{SeffVarianceAppA}) once the $x(\nu)=0$ approximation is made. 
Nevertheless, the series expansion proves to be useful to discuss the differences of the corrections in (\ref{cumulantMF}) and (\ref{cumulantBest}).

In Fig.~(\ref{Figure4}) we compare the results with DiagMC values obtained at small to intermediate coupling strengths \cite{Vlietinck2015} at a cutoff of $\Lambda=2000 \xi^{-1}$. 
We see that a significant correction to Feynman's original model is obtained by using the general memory kernel method, but nevertheless in the challenging intermediate coupling regime noticeable discrepancies remain. 
The corrected energy to second order discussed in this section yields excellent agreement with DiagMC in this regime. 
It should be emphasized that both the RG \cite{GrusdtRGFrohlich} and CGW \cite{ShchadilovaGaussianFrohlich} methods yield equally good agreement with DiagMC here. 

Finally, in Fig.~(\ref{Figure5}) we show how the logarithmic divergence observed in DiagMC can be completely retrieved in the corrected energy.
Once again, while the general memory kernel approach yields significant improvements to the original model system, the corrected energy is necessary to obtain further agreement with DiagMC. However, it should be noted that in this regime at $\alpha \approx 3$, especially at small cutoff values, we leave the weak to intermediate coupling regime and the approximation (\ref{cumulantBestMf}) can no longer be safely justified to accurately represent the second order cumulant correction. 
Nevertheless, the expression appears to be in excellent agreement with DiagMC, but we leave open the possibility that the exact second order correction would slightly overcorrect DiagMC in this regime, only to be brought back in the third order cumulant. 

\section{Conclusion}

In conclusion, in this work we explored extensions of Feynman's variational path integral treatment of the Bogoliubov-Fr\"{o}hlich model and addressed the issues of this method that were brought up in a number of works \cite{Vlietinck2015,GrusdtRGFrohlich,ShchadilovaGaussianFrohlich}. 
We show that two adjustments can be made to obtain major improvements to the original approach to this model that was first studied in \cite{Tempere2009Feynman}. 

First, instead of considering a coupled oscillator for the model action, a general quadratic action functional with a variational memory kernel is proposed. 
This method has already been studied for the original Fr\"{o}hlich model \cite{Adamkowski,Rosenfelder} but was found to yield only minor corrections for the energy. 
We show that this step is absolutely necessary to treat the Bogoliubov-Fr\"{o}hlich model, and obtain relatively good agreement with DiagMC at strong coupling.

To capture the difficult intermediate regime where the phonons of the model are strongly correlated \cite{ShchadilovaGaussianFrohlich}, even with this improvement noticeable discrepancies remain. 
For this reason we propose to include higher order corrections to the energy 
beyond the first order variational inequality, expanded around the general 
model action functional. 
These corrections have also been studied in the context of the original Fr\"{o}hlich model \cite{Mills,Marshall,SecondOrderCorrection}, but the studies were situated strictly within Feynman's approach and in addition the corrections were found to be small. 
In this work we have generalized previous results to the general memory kernel case and applied it to the Bogoliubov-Fr\"{o}hlich model. 
To obtain an easy semi-analytic expression for the correction we have proposed an approximation that naturally presents itself within the general memory kernel treatment. 
We estimated this approximation to be accurate in the weak to intermediate coupling regime and obtain excellent agreement with DiagMC. In addition, the correct logarithmic divergence of the model is retrieved. Renormalization procedures of the divergence are discussed in \cite{GrusdtRGFrohlich,Lampart}

This approach could be extended to many particles or to finite temperatures, which could be a way to probe the effect of thermal fluctuations on a system where quantum fluctuations are of great importance. 
Having seen how the second order correction around the optimal quadratic  action functional can be approximated by subtracting a single term from the perturbative correction with respect to a free particle, it would also be interesting to explore this in the context of higher order corrections. 

\begin{acknowledgments}
We gratefully acknowledge fruitful discussions with F. Brosens, S. N. Klimin, M. Houtput and S. Van Loon. We also acknowledge S. N. Klimin for referring us to the general memory kernel method for the Fr\"{o}hlich model.
T.I. acknowledges the support of 
the Research Foundation-Flanders (FWO-Vlaanderen) 
through the PhD Fellowship Fundamental Research, Project No. 1135521N. 
We also acknowledge financial support from the Research Foundation-Flanders (FWO-Vlaanderen) Grant No. G.0618.20.N, and  from the research council of the University of Antwerp.

\end{acknowledgments}

 \appendix

  \onecolumngrid 

 \section{Simplifying the second order correction}\label{AppendixA}

Written out in its full form, the effective action is given by:
\begin{equation}
\mathcal{\tilde{S}}_{\textrm{eff}}= - \frac{1}{8\pi }  \frac{  \alpha}{4 \pi\mu^2}   \int \mathbf{dk} V_{\mathbf{k}}^2 \int_0^{\beta} d\tau \int_0^{\beta} d\sigma  \mathcal{G}_{\mathbf{k}}(\tau-\sigma) e^{i \mathbf{k} \cdot \left[ \mathbf{r}(\tau)-\mathbf{r}(\sigma)\right]} . 
\end{equation}
The expectation value of the effective action with respect to the model system can be written in terms of 
\begin{equation}
 \mathcal{F}_{\mathbf{k}}(\tau-\sigma) = \expval{ e^{i \mathbf{k} \cdot \left[ \mathbf{r}(\tau)-\mathbf{r}(\sigma)\right]} },
 \label{Fdiscrete}
\end{equation}
as
\begin{equation}
\expval{\mathcal{\tilde{S}}_{\textrm{eff}}}= - \frac{  \alpha}{4 \pi\mu^2}  \frac{1}{2 }  \int_0^{\Lambda} dk~k^2 V_{\mathbf{k}}^2 \int_0^{\beta} d\tau \int_0^{\beta} d\sigma  \mathcal{G}_{\mathbf{k}}(\tau-\sigma) \mathcal{F}_{\mathbf{k}}(\tau-\sigma)
\label{SeffAppendix1}.
\end{equation}
Both $\mathcal{G}_{\mathbf{k}}(\tau-\sigma)$ and 
$\mathcal{F}_{\mathbf{k}}(\tau-\sigma)$ 
only depend on the difference $|\tau-\sigma|$ and are in addition $\beta$-periodic. 
This allows us to simplify the expectation value of the effective action to:
\begin{equation}
\frac{1}{\beta} \expval{\mathcal{\tilde{S}}_{\textrm{eff}}}= - \frac{  \alpha}{4 \pi\mu^2}   \int_0^{\Lambda} dk~k^2 V_{\mathbf{k}}^2 \int_0^{\beta/2} du \mathcal{G}_{\mathbf{k}}(u) \mathcal{F}_{\mathbf{k}}(u).
\end{equation}
As already seen in Sec.~(\ref{BestQuadraticAction}), for a general model action
$\mathcal{F}_{\mathbf{k}}(u)$ is given by:
\begin{equation}
 \mathcal{F}_{\mathbf{k}}(u) = \exp \left( - \frac{2k^2}{ \beta} \sum_{n=1}^{\infty} \frac{1- \cos(\nu_n u) }{x_n + \nu_n^2} \right).
 \label{Fdiscrete}
\end{equation}
The terms in the cumulant expansion (\ref{cumulant1}) can be derived using the $\lambda$-trick that has also been used in Sec.~(\ref{BestQuadraticAction}). 
It is not difficult to show that if a scaling parameter $x \rightarrow \lambda x$ is introduced in the memory kernel, the last two terms of (\ref{cumulant1}) can be written as:
\begin{align*}
&   \frac{1}{2\beta} \left( \expval{\mathcal{\tilde{S}}_{0}^2} - \expval{\mathcal{\tilde{S}}_{0}}^2   \right) = - \frac{1}{2} \left. \frac{\partial^2 F_0^{(\lambda)}}{\partial \lambda^2} \right|_{\lambda=1}, \\
&  \frac{1}{ \beta} \left( \expval{\mathcal{\tilde{S}}_{\textrm{eff}} \mathcal{\tilde{S}}_{0}}- \expval{\mathcal{\tilde{S}}_{\textrm{eff}}} \expval{\mathcal{\tilde{S}}_{0}}  \right) = -   \frac{1}{ \beta} \left. \frac{\partial \expval{\mathcal{\tilde{S}}_{\textrm{eff}}}_{\lambda}}{\partial \lambda} \right|_{\lambda=1}.
\end{align*}
The expression for $F_0^{(\lambda)}$ is given by (\ref{F_0}) and hence: 
\begin{equation}
\frac{1}{2\beta} \left( \expval{\mathcal{\tilde{S}}_{0}^2} - \expval{\mathcal{\tilde{S}}_{0}}^2   \right) = \frac{3}{2\beta} \sum_{n=1}^{\infty} \frac{x_n^2}{\left( \nu_n^2 + x_n \right)^2} .
\end{equation}
Similarly, $\expval{\mathcal{\tilde{S}}_{\textrm{eff}}}$ is given by Expression~(\ref{SeffAppendix1}). 
To include the $\lambda$-dependence, $x_n$ is substituted by $\lambda x_n$ in the memory function $\mathcal{F}_{\mathbf{k}}(u)$ after which the derivative can be taken. 
This yields:
\begin{align}
& \frac{1}{ \beta} \left( \expval{\mathcal{\tilde{S}}_{\textrm{eff}} \mathcal{\tilde{S}}_{0}}- \expval{\mathcal{\tilde{S}}_{\textrm{eff}}} \expval{\mathcal{\tilde{S}}_{0}}  \right)  =   \frac{4}{  \beta} \sum_{n=1}  \frac{x_n}{(\nu_n^2+ x_n)^2} \nonumber   \\
& \times \frac{\alpha}{4\pi \mu^2}  \int_0^{\Lambda} dk ~k^4 V_{\mathbf{k}}^2 \int_0^{\beta/2}  \sin(\frac{\nu_nu}{2})^2  \mathcal{G}_{\mathbf{k}}(u) \mathcal{F}_{\mathbf{k}}(u)  du . \label{expSeff2}
\end{align}
We can now recognize in (\ref{expSeff2}) the right-hand side 
of the iterative equation (\ref{xnu}). 
This means that if we are considering a perturbative correction on top of the memory kernel that solves (\ref{xnu}), we can write:
\begin{align}
& \frac{1}{ \beta} \left( \expval{\mathcal{\tilde{S}}_{\textrm{eff}} \mathcal{\tilde{S}}_{0}}- \expval{\mathcal{\tilde{S}}_{\textrm{eff}}} \expval{\mathcal{\tilde{S}}_{0}}  \right)  =   \frac{3}{  \beta} \sum_{n=1}  \frac{x_n^2}{(\nu_n^2+ x_n)^2}  \label{expSeff3}
\end{align}
which yields for the full second order correction around the optimized model action:
\begin{align}
 & \frac{1}{2\beta} \expval{ \left( \Delta \mathcal{S}- \expval{\Delta \mathcal{S}}  \right)^2}  =  \frac{1}{2\beta}  \left( \expval{\mathcal{\tilde{S}}_{\textrm{eff}}^2} - \expval{\mathcal{\tilde{S}}_{\textrm{eff}}}^2   \right)  -\frac{3}{2\beta} \sum_{n=1}^{\infty} \frac{x_n^2}{\left( \nu_n^2 + x_n \right)^2}.
  \label{Seff4}
\end{align}
Next, consider the variance of the effective action 
in the first square bracket of (\ref{Seff4}). 
The first term of the variance can be written as:
\begin{align}
\expval{\mathcal{\tilde{S}}_{\textrm{eff}}^2} = & ~ \frac{\pi^2}{(2\pi)^6} \left( \frac{\alpha}{4\mu^2} \right)^2 \int \mathbf{dk} \int \mathbf{ds}  V_{\mathbf{k}}^2 V_{\mathbf{s}}^2 \int_0^{\beta} d\tau_1 \int_0^{\beta} d\sigma_1 \int_0^{\beta} d\tau_2 \int_0^{\beta} d\sigma_2~ \mathcal{G}_{\mathbf{k}}(\tau_1-\sigma_1) \mathcal{G}_{\mathbf{s}}(\tau_2-\sigma_2) \nonumber \\
&  \times \expval{  e^{i \mathbf{k} \cdot \left[ \mathbf{r}(\tau_1)-\mathbf{r}(\sigma_1) \right]  + i \mathbf{s} \cdot  \left[ \mathbf{r}(\tau_2)- \mathbf{r}(\sigma_2)\right]}  } . 
\label{Seff5}
\end{align}
The generating function result (\ref{generatingfunction}) can now be used to find:

\begin{align}
&\expval{  e^{i \mathbf{k} \cdot \left[ \mathbf{r}(\tau_1)-\mathbf{r}(\sigma_1) \right]  + i \mathbf{s} \cdot  \left[ \mathbf{r}(\tau_2)- \mathbf{r}(\sigma_2)\right]}  } = \mathcal{F}_{\mathbf{k}}(\tau_1-\sigma_1) \mathcal{F}_{\mathbf{s}}(\tau_2-\sigma_2) \nonumber \\
& \times \exp \left(- \frac{\mathbf{k} \cdot \mathbf{s}}{4}  \zeta \left( \tau_1-\sigma_1, \tau_2-\sigma_2, \frac{\tau_1 + \sigma_1 - \tau_2 - \sigma_2}{2} \right) \right),
\end{align}
where $\zeta$ is given by:
\begin{equation}
\zeta(u_1,u_2,s) = \frac{32}{\beta} \sum_{n=1}^{\infty} \frac{\sin(\frac{\nu_n u_1}{2}) \sin(\frac{\nu_n u_2}{2}) \cos(\nu_n s) }{\nu_n^2+x_n}.
\label{zeta}
\end{equation}
The angle between $\mathbf{k}$ and $\mathbf{s}$ can be integrated out in (\ref{Seff5}) immediately. 
In addition we can see that the imaginary time integrals in (\ref{Seff5}) contain four variables, whereas the integrand only depends on $\tau_1-\sigma_1$, $\tau_2-\sigma_2$ and $ \left( \frac{\tau_1 + \sigma_1 - \tau_2 - \sigma_2}{2} \right)$. 
This allows to remove one integration variable and through the use of symmetry in the limit $\beta \rightarrow \infty$ significantly simplify the integral in similar spirit to what is done in \cite{SecondOrderCorrection}. 
Note however that even when divided by $\beta$, the integral (\ref{Seff5}) will still contain a divergence as $\beta \rightarrow \infty$ which is exactly canceled by subtracting its mean squared. 
Therefore in the limit of $\beta \rightarrow \infty$ we take both (\ref{Seff5}) and (\ref{SeffAppendix1}) together and obtain:
\begin{align}
& \frac{1}{2\beta} \left( \expval{\mathcal{\tilde{S}}_{\textrm{eff}}^2} -\expval{\mathcal{\tilde{S}}_{\textrm{eff}} }^2 \right) =\left( \frac{\alpha}{4\pi \mu^2} \right)^2   \int_0^{\Lambda} dk k^2 \int_0^{\Lambda} ds s^2  V_{\mathbf{k}}^2 V_{\mathbf{s}}^2 \nonumber \\
& \times   \int_0^{\beta/2} du_1   \int_0^{\beta/2} du_2   \mathcal{G}_{\mathbf{k}}(u_1) \mathcal{G}_{\mathbf{s}}(u_2) \mathcal{F}_{\mathbf{k}}(u_1) \mathcal{F}_{\mathbf{s}}(u_2) \int_0^{\beta/2} dz \left( \frac{\sinh \left[ \frac{ks}{4  } \zeta (u_1,u_2,z ) \right]}{ \frac{ks}{4  } \zeta (u_1,u_2,z )} - 1  \right).
\label{SeffVarianceAppA}
\end{align}
Contrary to the expression in \cite{Marshall,SecondOrderCorrection}, the quantity in the inner integral is a $\sinh(x)/x$ function rather than an $\arcsin(x)/x$ function due to the fact that the momentum integrals cannot be performed analytically. It will prove to be useful to replace the hyperbolic sine function by its Taylor expansion:
\begin{equation}
\frac{\sinh(x)}{x} = 
\sum_{n=0}^{\infty} \frac{x^{2n}}{(2n+1)!}
\end{equation}
which yields
\begin{align}
& \frac{1}{2\beta} \left( \expval{\mathcal{\tilde{S}}_{\textrm{eff}}^2} -\expval{\mathcal{\tilde{S}}_{\textrm{eff}} }^2 \right) =\left( \frac{\alpha}{4\pi \mu^2} \right)^2   \int_0^{\Lambda} dk ~ k^2 \int_0^{\Lambda} ds ~ s^2  V_{\mathbf{k}}^2 V_{\mathbf{s}}^2 \nonumber \\
& \times   \int_0^{\beta/2} du_1   \int_0^{\beta/2} du_2   \mathcal{G}_{\mathbf{k}}(u_1) \mathcal{G}_{\mathbf{s}}(u_2) \mathcal{F}_{\mathbf{k}}(u_1) \mathcal{F}_{\mathbf{s}}(u_2) \int_0^{\beta/2} dz  \sum_{n=1}^{\infty}  \frac{1}{(2n+1)!} \left( \frac{ks}{4 } \zeta(u_1,u_2,z )\right)^{2n}.
\label{Svariance1}
\end{align}
Let us also define the individual terms of the sum, and emphasize their dependence on the memory kernel $x(\nu)$:
\begin{align}
& \sigma_n[x(\nu)] = \frac{1}{(2n+1)!} \left( \frac{\alpha}{4\pi \mu^2} \right)^2   \int_0^{\Lambda} dk k^2 \int_0^{\Lambda} ds s^2  V_{\mathbf{k}}^2 V_{\mathbf{s}}^2 \nonumber \\
& \times   \int_0^{\beta/2} du_1   \int_0^{\beta/2} du_2   \mathcal{G}_{\mathbf{k}}(u_1) \mathcal{G}_{\mathbf{s}}(u_2) \mathcal{F}_{\mathbf{k}}(u_1) \mathcal{F}_{\mathbf{s}}(u_2) \int_0^{\beta/2} dz  
\left( \frac{ks}{4 }  \zeta(u_1,u_2,z )\right)^{2n},
\label{sigman}
\end{align}
such that the entire second order cumulant is written as:
\begin{equation}
\frac{1}{2\beta} \expval{ \left( \Delta \mathcal{S}- \expval{\Delta \mathcal{S}}_0 \right)^2}_0  = \sum_{n=1}^{\infty} \sigma_n[x(\nu)] -\frac{3}{2\beta} \sum_{n=1}^{\infty} \frac{x_n^2}{\left( \nu_n^2 + x_n \right)^2}.
\label{sigmasum1}
\end{equation}
Note that the second term in (\ref{sigmasum1}) was obtained by assuming an expansion around the optimal memory kernel action and hence the same has to be done for the rest of the terms. 
Unfortunately, for a(n) (optimized) memory kernel $x(\nu)$ that has no trivial expression we cannot analytically perform the five-fold integral in (\ref{Svariance1}) or (\ref{sigman}), which is difficult even numerically. The exception to this is the $n=1$ expansion term:
\begin{align}
& \sigma_1[x(\nu)] = \frac{1}{6 (4 )^2} \left( \frac{\alpha}{4\pi \mu^2} \right)^2   \int_0^{\Lambda} dk k^4 \int_0^{\Lambda} ds s^4  V_{\mathbf{k}}^2 V_{\mathbf{s}}^2 \nonumber \\
& \times   \int_0^{\beta/2} du_1   \int_0^{\beta/2} du_2   \mathcal{G}_{\mathbf{k}}(u_1) \mathcal{G}_{\mathbf{s}}(u_2) \mathcal{F}_{\mathbf{k}}(u_1) \mathcal{F}_{\mathbf{s}}(u_2) \int_0^{\beta/2} dz       \zeta(u_1,u_2,z )^{2}.
\label{sigma1a}
\end{align}
By substituting $\zeta$ as given in (\ref{zeta}) and using the orthogonality of the cosine, the $z$ integral can be performed:
\begin{equation}
\int_0^{\beta/2} dz       \zeta(u_1,u_2,z )^{2} = \frac{32^2}{4 \beta} \sum_{n=1}^{\infty} \frac{\sin(\frac{\nu_n u_1}{2} )^2 \sin(\frac{\nu_n u_2}{2} )^2 }{(\nu_n^2+x_n)^2}.
\label{integralzeta}
\end{equation}
The remaining four-fold integral completely decouples in each term of the sum in (\ref{integralzeta}) and can be slightly simplified to:
\begin{align}
& \sigma_1[x(\nu)] =  \frac{3}{2 \beta } \sum_{n=1}^{\infty} \frac{1}{(\nu_n^2+x_n)^2}    \left[ \frac{\alpha}{3\pi \mu^2} \int_0^{\Lambda} dk k^4  V_{\mathbf{k}}^2  \int_0^{\beta/2} du   \mathcal{G}_{\mathbf{k}}(u)  \mathcal{F}_{\mathbf{k}}(u)  \sin(\frac{\nu_n u}{2} )^2 \right]^2.
\label{sigma1b}
\end{align}
The integral inside the square brackets is once again exactly the right-hand side of the iterative equation (\ref{xnu}) which means that for the optimal memory kernel:
 \begin{align}
& \sigma_1[x(\nu)] =  \frac{3}{2 \beta} \sum_{n=1}^{\infty} \frac{x_n^2}{(\nu_n^2+x_n)^2}     
\label{sigma1c}
\end{align}
cancels with the contribution from the other terms in (\ref{sigmasum1}). 
For an expansion around the optimal memory kernel the second order cumulant is written as:
\begin{equation}
\frac{1}{2\beta} \expval{ \left( \Delta \mathcal{S}- \expval{\Delta \mathcal{S}} \right)^2}  = \sum_{n=2}^{\infty} \sigma_n[x(\nu)]  .
\label{sigmasum2}
\end{equation}

\section{Calculating the approximated correction}\label{AppendixB}

In this appendix we will obtain a semi-analytic expression for the second order correction within the approximation discussed in Sec. (\ref{SecondOrder}):
\begin{align}
 & \frac{1}{2\beta} \expval{ \left( \Delta \mathcal{S}- \expval{\Delta \mathcal{S}}  \right)^2}^{(\textrm{approx.})} = \sum_{n=2}^{\infty} \sigma_{n}[0].
  \label{cumulantMFApp1}
\end{align}
The fact that the memory kernel vanishes, significantly simplifies the integral. 
First, expression (\ref{sigman}) is rewritten using the symmetry around $\beta/2$ to fold the $u_1$,$u_2$ integration domain in half:
\begin{align}
& \sigma_n[x(\nu)] = \frac{2}{4^{2n}}\frac{1}{(2n+1)!} \left( \frac{\alpha}{4\pi \mu^2} \right)^2   \int_0^{\Lambda} dk k^{2+2n} \int_0^{\Lambda} ds s^{2+2n}   V_{\mathbf{k}}^2 V_{\mathbf{s}}^2 \nonumber \\
& \times   \int_0^{\beta/2} du_1   \int_0^{u_1} du_2   \mathcal{G}_{\mathbf{k}}(u_1) \mathcal{G}_{\mathbf{s}}(u_2) \mathcal{F}_{\mathbf{k}}(u_1) \mathcal{F}_{\mathbf{s}}(u_2) \int_0^{\beta/2} dz      \zeta(u_1,u_2,z )^{2n}.
\label{sigmanAppB1}
\end{align}
Next, observe that for $x(\nu)=0$, the memory functions $\mathcal{F}_{\mathbf{k}}(u)$ simplify and in the limit of zero temperature this expression can be written as:
\begin{align}
& \sigma_n[0] = \frac{2}{4^{2n}}\frac{1}{(2n+1)!} \left( \frac{\alpha}{4\pi \mu^2} \right)^2   \int_0^{\Lambda} dk k^{2+2n} \int_0^{\Lambda} ds s^{2+2n}   V_{\mathbf{k}}^2 V_{\mathbf{s}}^2 \nonumber \\
& \times   \int_0^{\beta/2} du_1   \int_0^{u_1} du_2  e^{-a(k) u_1} e^{-a(s) u_2} \int_0^{\beta/2} dz      \zeta(u_1,u_2,z )^{2n},
\label{sigmanAppB2}
\end{align}
where the short hand notation with $\omega_k$ from (\ref{omegak}) is introduced:
\begin{equation}
a(\mathbf{k})= \omega_k + \frac{k^2}{2m}.
\end{equation}
As has already been observed in the weak-coupling limit of \cite{SecondOrderCorrection}, in the absence of a memory kernel the expression for $\zeta$ becomes quite simple (note that our $\zeta$ is defined differently but the same structure holds):
\begin{align*}
\zeta(u_1,u_2,z) = \begin{cases} 4u_2 \hspace{5pt} & \textrm{for} \hspace{5pt} z < \frac{u_1-u_2}{2}, \\
 2u_1 + 2u_2 - 4z \hspace{5pt} & \textrm{for} \hspace{5pt} \frac{u_1-u_2}{2} < z < \frac{u_1+u_2}{2}, \\
 0  \hspace{5pt} & \textrm{for} \hspace{5pt} \frac{u_1+u_2}{2} < z .
 \end{cases}
\end{align*}
The integral over $z$ can now be analytically performed:
\begin{align}
&  \int_0^{\beta/2} dz      \zeta(u_1,u_2,z )^{2n}= 4^{2n} u_2^{2n} \frac{u_1-u_2}{2} + 4^{2n} \frac{u_2^{2n+1}}{2n+1}.
\label{zetaintegral}
\end{align}
This allows us to write (\ref{sigmanAppB2}) as:
\begin{align}
& \sigma_n[0] =  \frac{2}{(2n+1)!} \left( \frac{\alpha}{4\pi \mu^2} \right)^2   \int_0^{\Lambda} dk k^{2+2n} \int_0^{\Lambda} ds s^{2+2n}   V_{\mathbf{k}}^2 V_{\mathbf{s}}^2 \nonumber \\
& \times   \int_0^{\beta/2} du_1   \int_0^{u_1} du_2  e^{-a(k) u_1} e^{-a(s) u_2}   \left[ u_2^{2n} \frac{u_1-u_2}{2} +  \frac{u_2^{2n+1}}{2n+1}\right].
\label{sigmanAppB3}
\end{align}
The integrals over $u_1$ and $u_2$ are given by
\begin{equation}
\int_0^{\infty} du_1   \int_0^{u_1} du_2  e^{-a(\mathbf{k}) u_1} e^{-a(\mathbf{s}) u_2}   \left[ u_2^{2n} \frac{u_1-u_2}{2} +  \frac{u_2^{2n+1}}{2n+1}\right]= \frac{\left(3 a(k) + a(s) \right)  n \Gamma(2n) }{a(k)^2 \left( a(k) + a(s) \right)^{2+2n}}.
\end{equation}
Since $n$ is an integer $n \Gamma(2n) =  (2n)!/2$ and therefore:
\begin{align}
& \sigma_n[0] =  \frac{1}{(2n+1)} \left( \frac{\alpha}{4\pi \mu^2} \right)^2   \int_0^{\Lambda} dk  \int_0^{\Lambda} ds     V_{\mathbf{k}}^2 V_{\mathbf{s}}^2  \frac{k^{2+2n} s^{2+2n}\left(3 a(k) + a(s) \right)   }{a(k)^2 \left( a(k) + a(s) \right)^{2+2n}}.
\label{sigmanAppB3}
\end{align}
Finally, we can define
\begin{equation}
Q = \sum_{n=2}^{\infty} \frac{1}{(2n+1)} \left( \frac{1}{4\pi \mu^2} \right)^2   \int_0^{\Lambda} dk  \int_0^{\Lambda} ds     V_{\mathbf{k}}^2 V_{\mathbf{s}}^2  \frac{k^{2+2n} s^{2+2n}\left(3 a(k) + a(s) \right)   }{a(k)^2 \left( a(k) + a(s) \right)^{2+2n}},
\end{equation}
such that the full approximate second order correction is given by:
\begin{equation}
E_{2} = - \alpha^2 Q.
\end{equation}

   \twocolumngrid 
\end{document}